\documentclass[prl, amsmath, showpacs, twocolumn, sort&compress, floatfix, amssymb, noeprint, longbibliography]{revtex4-1}

\usepackage{graphicx}
\usepackage{dcolumn}
\usepackage{bm}
\usepackage[dvipsnames]{xcolor}
\usepackage{hyperref}
\usepackage{siunitx}

\begin{document}

\title{Control of light-atom solitons and atomic transport by optical vortex beams propagating through a Bose-Einstein Condensate}

\author{Grant Henderson}
 \email{grant.henderson@strath.ac.uk}
\author{Gordon R. M. Robb}
\author{Gian-Luca Oppo}
\author{Alison M. Yao}
\affiliation{SUPA \& Department of Physics, University of Strathclyde, Glasgow, Scotland, UK}

\date{\today}

\begin{abstract}
We model propagation of far-red-detuned optical vortex beams through a Bose-Einstein Condensate using nonlinear Schrödinger and Gross-Pitaevskii equations. We show the formation of coupled light/atomic solitons that rotate azimuthally before moving off tangentially, carrying angular momentum.  The number, and velocity, of solitons, depends on the orbital angular momentum of the optical field. Using a Bessel-Gauss beam increases radial confinement so that solitons can rotate with fixed azimuthal velocity. Our model provides a highly controllable method of channelling a BEC and atomic transport.
\end{abstract}

\maketitle

Solitons are localized fields that maintain their spatial profile as they propagate. They have been investigated and realized in fields as diverse as optical fibres \cite{PhysRevLett.45.1095}, hydrodynamics \cite{PhysRevLett.52.1421}, ferromagnetic and anti-ferromagnetic systems \cite{KOSEVICH1990117}, superconductors \cite{PhysRevLett.88.017002} and even cosmology \cite{Dauxois_2010}. Bright \cite{Khaykovich1290}, dark \cite{denschlag2000generating} and lattice \cite{PhysRevLett.92.230401} solitons have also been observed in Bose-Einstein Condensates (BECs). 
In nonlinear optics, spatial optical solitons \cite{Dauxois_2010} arise when the diffraction of a Gaussian beam is carefully balanced by self-focusing due to a Kerr nonlinear medium. However, when the optical field carries orbital angular momentum (OAM) \cite{Allen92}, it fragments into solitons, with the number of formed solitons depending, generally, on the OAM index, $m$ \cite{PhysRevLett.79.2450,PhysRevLett.87.033901}. This has been confirmed experimentally using hot sodium \cite{Bigelow04} and rubidium vapours \cite{PhysRevLett.117.233903} as the Kerr medium. Similar fragmentation has been seen in nonlinear colloidal suspensions \cite{Walasik17, Sun18}.

In this letter we use coupled nonlinear Gross-Pitaevskii and Schrödinger equations to describe the propagation of far-detuned optical fields through a cigar-shaped BEC. 
We start by confirming that, for weakly repulsive atomic interactions, our model captures the formation of coincident patterns seen in \cite{Saffman2001} for light that is red-detuned with respect to the atoms (so that atoms are attracted to intensity peaks).

We then show that if the light carries OAM, it fragments into solitons during propagation, suggesting that the BEC is behaving like an effective Kerr superfluid. As the atoms are attracted to intensity peaks they are ``captured'' by the optical solitons, resulting in coupled light-atom solitons. We show that both the optical and atomic solitons carry angular momentum and that the number of solitons formed, and their velocities, is dependent on the OAM of the optical field. The radial spread of the solitons can be reduced by replacing the Laguerre-Gauss optical field with a Bessel-Gauss field.

Our results suggest a highly effective means of channelling large BEC transverse distributions into a given number of tightly confined solitons, presenting a novel method of controllable atomic transport. 

\begin{figure}[tbp]
    \includegraphics[width=8.6cm]{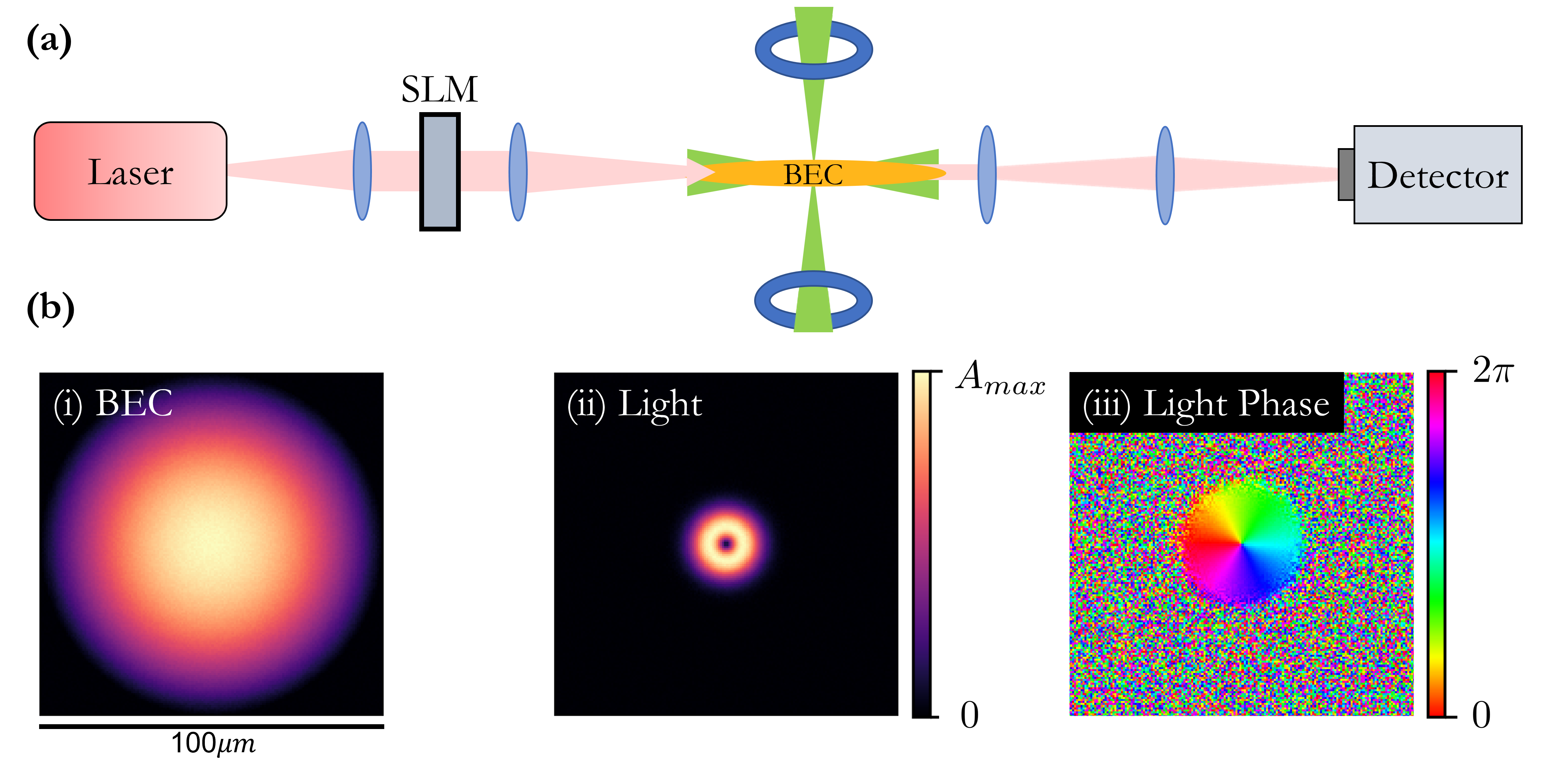}
    \caption{(a) An input laser beam is incident on an SLM, which can add OAM, before propagating through a cigar-shaped BEC medium, suspended by additional horizontal and vertical trapping fields, to a detector. (b) Transverse cross-sections of Thomas-Fermi BEC amplitude with $w_{\psi} = 50.0 \mu$m (i), and Laguerre-Gauss optical field amplitude (ii) and phase (iii) with $m = 1$ and $w_F = 10 \mu$m.}
    \label{fig:ExpSetup}
\end{figure}

A schematic of the proposed set-up is shown in Fig \ref{fig:ExpSetup} (a). A coherent Gaussian beam of waist $w_F$ and wavelength $\lambda$, typically from a diode laser, is incident on a spatial light modulator (SLM) with an ``m"-forked diffraction grating \cite{Heckenberg92} which can convert it to an optical vortex beam carrying OAM of $m \hbar$ per photon \cite{Allen92}. This optical field then propagates through a cigar-shaped BEC, that is suspended by additional horizontal and vertical trapping fields, before being focused onto a detector.

Our numerical model is similar to that of \cite{Saffman2001}, but includes terms in $L_3$ and $\sigma_{sat}$ describing three-body loss and optical saturation, respectively:
\begin{eqnarray}
 \partial_\zeta \psi &=& i \nabla_\perp ^2 \psi - i  \left ( s |F|^2 + \beta_{col} |\psi|^2 - iL_3|\psi|^4 \right )\psi \label{eqn:atoms}, \\
 \partial_\zeta F  &=& i \nabla_\perp ^2 F + i \left (\frac{ -s |\psi|^2}{1 + \sigma_{sat}|F|^2} \right ) F \label{eqn:light_coupling} .
\end{eqnarray}
Here $\psi = \psi(\xi,\eta, \zeta)$ and $F = F(\xi,\eta,\zeta)$ are the slowly varying amplitudes of the BEC wavefunction and optical field, respectively, with wavenumbers $k_a =  m_a v_a/\hbar$ and $k_L = 2 \pi / \lambda$. $m_a$ is the atomic mass, $v_a$ the velocity of the atomic beam, and we assume that 
$k_a \approx k_L$. The transverse and longitudinal dimensions are scaled according to $(\xi,\eta) = \sqrt{2}(x,y)/w_F$ and $\zeta = z/(2 z_R)$, where $z_R = \pi w_F^2/\lambda$ is the Rayleigh range.

Eq. (\ref{eqn:atoms}), which describes the atomic dynamics, is a reduction of the 3D Gross-Pitaevskii equation (GPE) to 2D \cite{PhysRevA.65.043614}. A more accurate reduction from 3D to 2D is given by the non-polynomial Schr\"{o}dinger equation (NPSE), which includes additional nonlinear terms to allow for variations in the axial width of the atomic cloud \cite{PhysRevA.65.043614,PhysRevA.79.053620}. We have verified that for our simulations both models result in identical dynamics and so, for ease of analysis, we will proceed using the 2D GPE (\ref{eqn:atoms}).

In Eq. (\ref{eqn:atoms}) the transverse Laplacian term ($\nabla_\perp ^2$) represents the kinetic energy contributions and the $s |F|^2 \psi$ term describes light-induced focusing/defocusing due to the dipole interaction. As defined in \cite{Saffman2001}, $\beta_{col}$ is directly proportional to the interatomic scattering length of interactions of ground-state atoms, $a_{gg}$, and thus the term $\beta_{col} |\psi|^2\psi$ describes attractive or repulsive interactions depending on the sign of $a_{gg}$.
Typical scattering parameter values are given in Table \ref{tab:ScatLens}, using atomic parameters from \footnote{D. A. Steck, Alkali D Line Data, available online at \url{http://steck.us/alkalidata} (2022)}. 
For clarity we have chosen BEC parameters to match those found for a BEC of weakly {\em repulsive} Caesium atoms ($a_{gg} = 15.7 a_0$, with $a_0$ the Bohr radius), giving $\beta_{col} = 3.5$, but we emphasise that our analysis is applicable over a wide range of scattering lengths, accessible around the Feshbach resonance \cite{PhysRevLett.125.183602}.
As mentioned, we also employ a term $L_3|\psi|^4 \psi$ describing three-body loss ($L_3 \approx 10^{-4}$) to arrive at an accurate description of the evolution of the BEC wave-function (matter-wave) in high-density regimes.
The selected value of $L_3$ is in agreement with estimations for Caesium \cite{kraemer2006evidence,PhysRevLett.125.183602}.

\begin{table}[tbp]
    \centering
    \begin{ruledtabular}
    \begin{tabular}{c c c}
    Species & $a_{gg} [\mathrm{Bohr \, radius, \,} a_0]$ & $\beta_{col}$ \\ \colrule
    Lithium & -27.6 & -8.22 \\ 
    Sodium & 260 & 117 \\ 
    Rubidium (87) & 110 & 21.6 \\ 
    Caesium & $-500 \rightarrow 500$ & $-110 \rightarrow 110$
    \end{tabular}
    \caption{Typical ground state scattering parameters $a_{gg}$ of various BEC species with corresponding $\beta_{col}$ values.}
    \label{tab:ScatLens}
    \end{ruledtabular}
\end{table}

Eq. (\ref{eqn:light_coupling}) describes the dynamics of the optical field. Here the transverse Laplacian, $\nabla_\perp ^2$, describes diffraction, the term $s |\psi|^2 F$ describes a focusing/defocusing proportional to the atomic density, and $\sigma_{sat}$ describes optical saturation, which is critical to prevent soliton collapse in two transverse dimensions in the case of pure Kerr media \cite{Berge98}. In the Kerr case $\sigma_{sat} = (4 P_L) / (3 I_{sat} w_F^2)$, where $P_L$ is the power of the incident laser beam and $I_{sat}$ is the saturation intensity \cite{PhysRevLett.117.233903}. For typical parameter values we find $\sigma_{sat} \approx 10^{-3}$ \cite{PhysRevLett.79.2450}. 
Finally we note that higher order terms corresponding to dipole-dipole forces have been neglected since they only marginally affect the system dynamics.

In both (\ref{eqn:atoms}) \& (\ref{eqn:light_coupling}) the parameter $s =\pm 1$ provides a control for the nature of the BEC-optical field dipole coupling. For $s = +1$, the optical field is blue-detuned, and the BEC can be described as `dark-seeking' with relation to the optical field. For $s = -1$, the optical field is red-detuned, and the BEC can be described as `light-seeking', and behaves like a self-focusing medium \cite{Saffman2001}.
We numerically integrate Eq. (\ref{eqn:atoms}) and (\ref{eqn:light_coupling}) using a split-step Fourier method, and include noise at 1\% of the amplitude on the initial fields.

The initial wave-function of the BEC is a Thomas-Fermi (TF) distribution of amplitude $A_{\psi}$ and transverse width $w_{\psi}$ with any negative values set to zero:
\begin{equation}
\psi(\xi,\eta, \zeta(0)) = A_{\psi} \left [ 1 - \left (\xi^2 + \eta^2 \right )/2 w_{\psi}^2  \right ].
\label{eqn:AtomicIC}
\end{equation}
Here we choose $w_{\psi} = 50.0 \mu$m, corresponding to an experimentally realisable BEC with a transverse diameter of $100 \mu $m, and we assume a longitudinal length of around $2$mm \cite{Carli_2019,PhysRevLett.87.050404}.
We consider an optical field with wavelength $\lambda=720$ nm and initial Laguerre-Gaussian profile of amplitude $A_F$ and OAM $m$ at the beam waist $w_F$ \cite{Barnett07}:
\begin{eqnarray}
F(\xi, \eta, \zeta(0)) &=&  A_F \hspace{0.5mm} LG_0^m \left(\xi, \eta \right )/\mathrm{max}|LG_0^m| , \label{eqn:OpticalIC} \\
\mathrm{where} \hspace{1.0mm} LG_0^m \left(\xi, \eta \right )  &=& (\xi^2 + \eta^2)^{|m|} e^{-\frac{(\xi^2 + \eta^2)}{2}} e^{i m \varphi} . \hspace{0.5cm}
\label{eqn:OAMmode}
\end{eqnarray}
Fig. \ref{fig:ExpSetup} (b) shows transverse cross-sections of typical initial fields ($\zeta = 0)$. Panel (i) shows the amplitude of the Thomas-Fermi BEC with $w_{\psi} = 50.0 \mu$m. Panels (ii) \& (iii) show the amplitude and phase, respectively, of a Laguerre-Gauss optical field with $m = 1$ and beam waist $w_F = 10 \mu$m chosen so that the beam propagates for several Rayleigh ranges inside the atomic medium ($z_R \approx 0.44$ mm). 

\begin{figure}[tbp]
    \includegraphics[width=8.6cm]{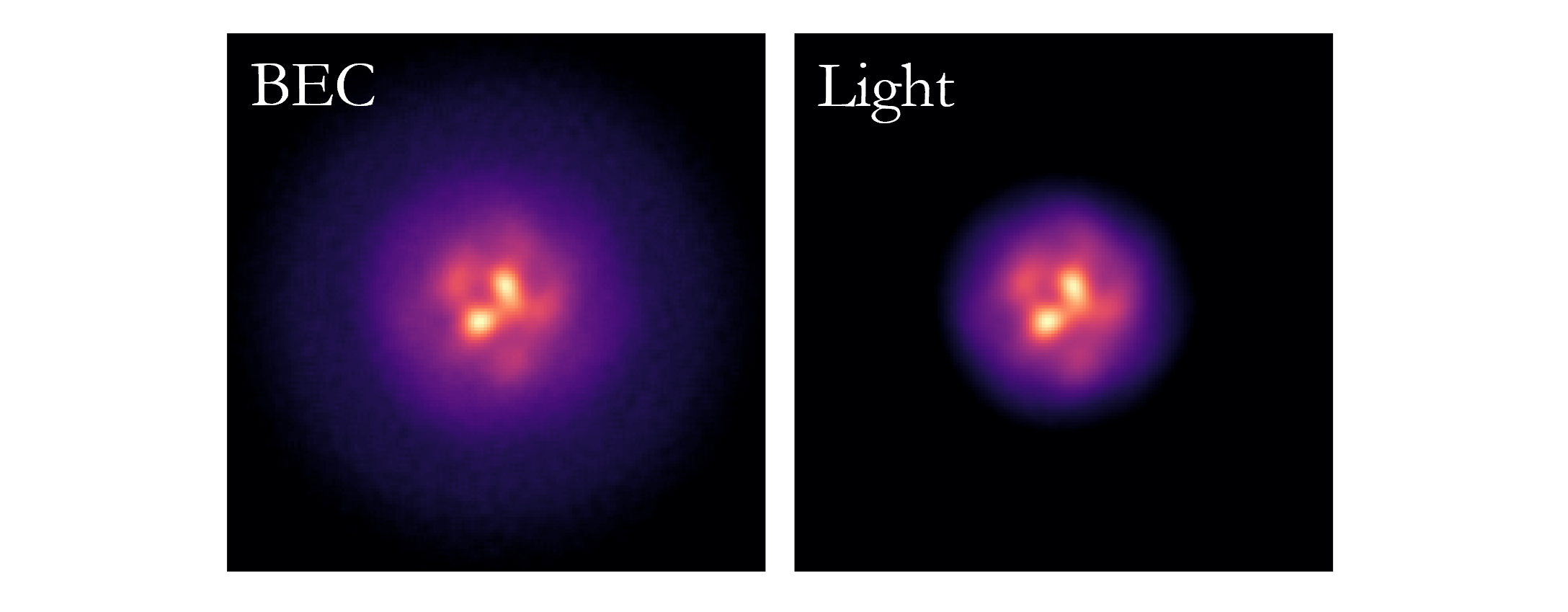}
    \caption{Coincident pattern formation in both BEC and optical fields for Gaussian input profiles as in Ref. \cite{Saffman2001}. 
    Transverse scales as in Fig. \ref{fig:ExpSetup}. Parameters: $\beta_{col} = 3.5, L_3 = 0.00022, s = -1, w_F = 50 \mu$m, $A_F = 6, w_{\psi} = 50 \mu$m, $A_{\psi} = 6$.}
    \label{fig:Patterns}
\end{figure}

We start by confirming that our model accurately reproduces the results shown in \cite{Saffman2001,PhysRevLett.81.65} for the red-detuned case, $s = -1$. To maximise the area in which the patterns can form, the initial optical field is a Gaussian ($m = 0$) with the same beam waist, $w_F = 50 \mu$m, as the BEC. The normalised field amplitudes are $A_F = 6$ and $A_{\psi} = 6$, corresponding to input powers on the order of mW and a total atom number of $\sim 10^5$, respectively.
Fig. \ref{fig:Patterns} shows the expected formation of {\em coincident} filament structures at $\zeta = z_R$ arising from a modulational instability due to the dipole interactions between the coupled BEC (left) and optical (right) fields \cite{Saffman2001}. Soon after the formation of filaments one observes on-axis collapse in both fields as the focusing nonlinearities overwhelm the system dynamics, akin to the BEC collapse experimentally studied in \cite{donley2001dynamics}.
Although outwith the scope of this letter, our model also confirms similar results seen in 1D \cite{PhysRevA.83.013608} that show that on-axis collapse can be avoided when the amplitude of the BEC is significantly different to that of the optical field.
In that case the dominant dynamics are linear rather than the highly nonlinear dynamics which we report here, with the optical field acting more as a potential on the BEC rather than a coupled field.

We now consider the effect of adding OAM to the optical beam.
With respect to our previous initial conditions, the main difference is that the optical field now has a $0 \rightarrow 2 m \pi$ azimuthal phase and the corresponding on-axis vortex produces a ring-like intensity profile, as shown in Fig. \ref{fig:ExpSetup} (b) for $m = 1$. 
We use $A_F = A_{\psi}=9.5$, and choose $w_F = 10\mu $m so that the beam propagates for several Rayleigh ranges inside the atomic medium ($z_R \approx 0.44$ mm), but emphasise that similar behaviour is obtained over a wide range of initial conditions. 

\begin{figure}[tbp]
    \includegraphics[width=8.6cm]{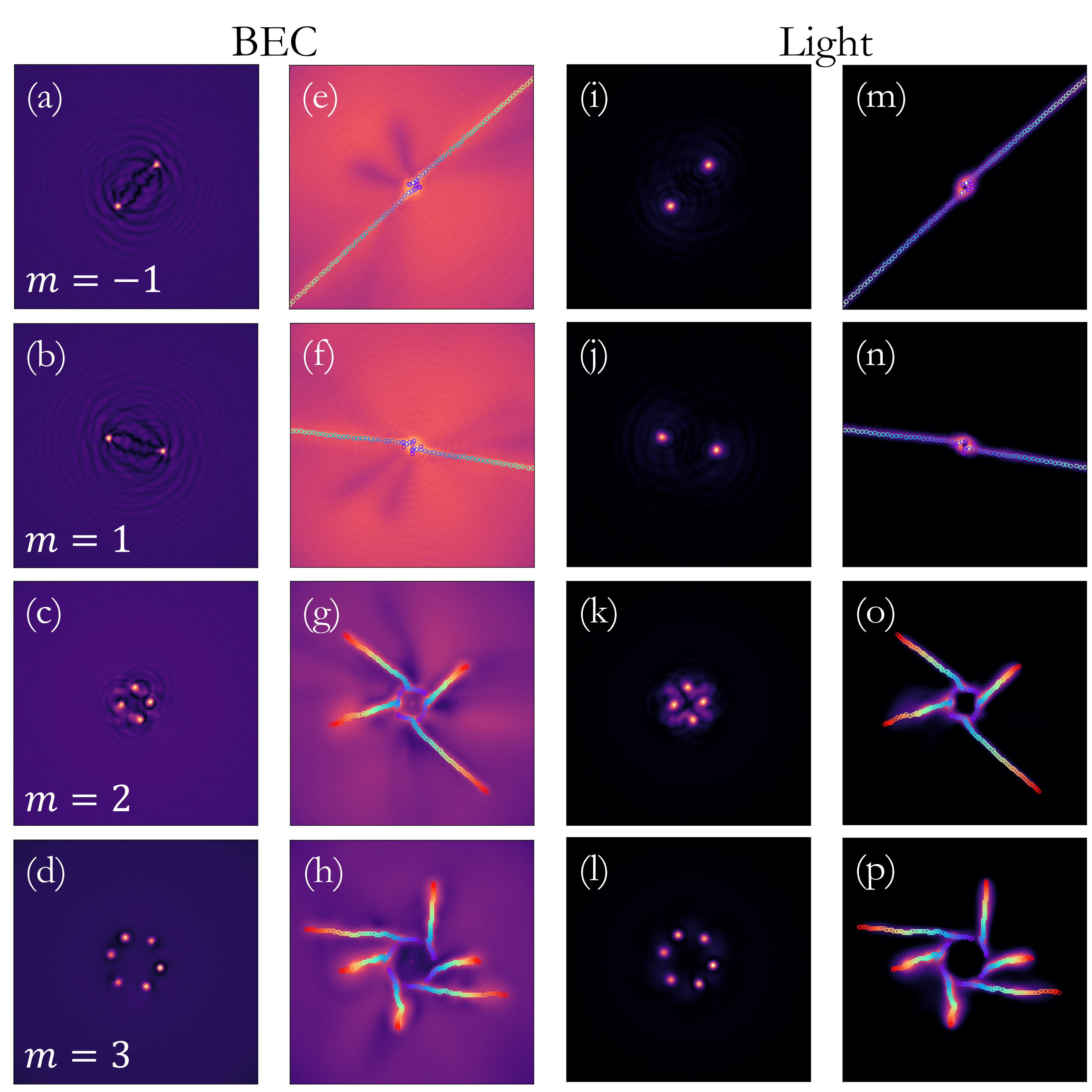}
    \caption{Panels (a)-(d) \& (i)-(l): Transverse amplitude cross-section of BEC and optical fields, respectively, for $m = -1, 1, 2, 3$ (top to bottom) at $\zeta = z_R$. Panels (e)-(h) \& (m)-(p): superimposed images of transverse BEC and optical amplitude distributions, respectively, $\zeta = 0.5 \rightarrow 4 z_R$. 
    Parameters as in Fig. \ref{fig:Patterns}, with $w_F = 10 \mu$m, $A_F = 9.5, A_{\psi} = 9.5$.}
    \label{fig:MainResults}
\end{figure}

Fig. \ref{fig:MainResults} shows the resultant optical and atomic fields after numerical integration of (\ref{eqn:atoms})-(\ref{eqn:light_coupling}) with optical beams carrying OAM of $m = -1, 1, 2, $ \& $3$.
We find that adding OAM has a profound effect on the dynamics: the light-seeking atoms now move radially towards the optical ring after which the dynamics of the BEC is closely coupled to that of the light and both fields start to form distinct solitons rather than narrow filaments, in spite of repulsive BEC interactions. Although both atomic density and light intensity increase significantly within these peaks, there is no collapse of the wave function even with negligible three-body loss. Moreover, we have verified that ring-shaped optical intensity profiles {\em without} the optical vortex, do undergo collapse. Panels (a)-(d) and (i)-(l) show the formation of these $2 |m|$ BEC and optical soliton peaks, respectively, at $\zeta = z_R$.

Initially the angular momentum leads to an azimuthal motion of the atomic peaks around the ring, analogous to persistent currents \cite{PhysRevLett.10.93}, and we find that the angular velocity of the solitons is inversely proportional to $m^2$. 
In general this ``current" lasts for around $0.75 z_R$ before diffractive dynamics begins to dominate and the peaks are ejected {\em tangentially} to the ring, thus carrying away the angular momentum. This suggests a means of realising atomic currents within a BEC over a wide range of longitudinal propagation distances as determined by the optical Rayleigh range.

In panels (e)-(h) and (m)-(p) we demonstrate the tangential motion of the solitons by overlaying a succession of transverse amplitude distributions from $\zeta = 0.5 z_R$ to $4 z_R$. We superimpose rainbow contours to highlight the propagation distance (blue at $\zeta = 0.5 z_R$, red at $\zeta = 4 z_R$). We find that the solitons move with a constant transverse velocity that is inversely proportional to $m$. This is particularly evident for the $m = -1$ and $m = 1$ cases where the solitons move in opposite directions and agrees very well with previous studies of fragmentation of OAM beams propagating in Kerr self-focusing media, predicted in \cite{PhysRevLett.79.2450,PhysRevLett.87.033901} and more recently demonstrated experimentally in \cite{PhysRevLett.117.233903}.

We are now in a second region of controllable atomic transport dynamics where simply by changing the OAM of the optical input field we can control both the number of atomic solitons formed and their tangential velocities. Again, the velocity is scaled by the Rayleigh range, meaning that it is possible to realise these controllable atomic transport dynamics across a wide range of longitudinal propagation distances, transverse field sizes and OAM values.

\begin{figure}[tbp]
    \includegraphics[width=8.6cm]{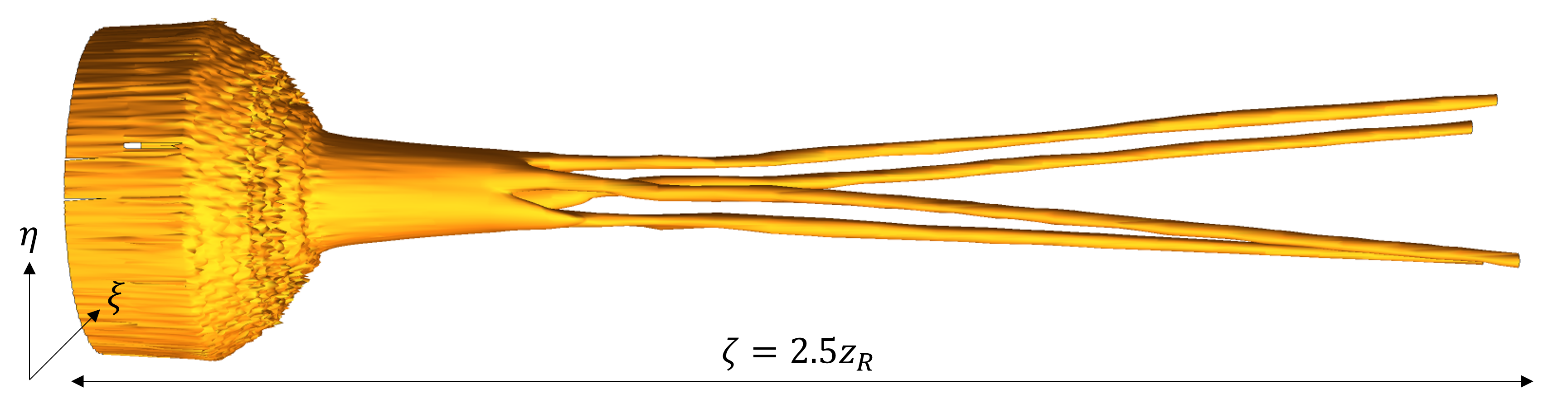}
    \caption{Three-dimensional BEC distribution for the $m = 2$ case of Fig. \ref{fig:MainResults}, $\zeta = 0 \rightarrow 2.5 z_R$. Transverse scales as in Fig. \ref{fig:ExpSetup}. Parameters as in Fig. \ref{fig:MainResults}.}
    \label{fig:3DProf}
\end{figure}

The overall behaviour of the system is summarised well in Fig. \ref{fig:3DProf}, which shows in 3D the re-distribution of the atoms as the far-red-detuned light propagates along the length of the BEC for the case of $m = 2$. The atoms, initially in a TF distribution, are focused onto a ring before splitting into four channels that {\em twist} as they propagate. 

We find that the coupled off-axis soliton formation process is robust across a wide range of OAM values, initial field amplitudes, beam sizes and BEC scattering parameters and both weakly attractive and repulsive interactions in the range $-20 a_0 < a_{gg} < 50 a_0$ corresponding to $-4 < \beta_{col} < 11$.
Notably, we find that both optical and atomic solitons propagate tangentially with little change to their shape or amplitude until they reach the transverse limits of the BEC.

We note that we can extend the duration of the azimuthal rotation and decrease the transverse motion of the solitons by replacing the Laguerre-Gauss mode (\ref{eqn:OAMmode}) with an equivalent Bessel-Gauss (BG) mode:
\begin{equation}
F_{BG}(\xi, \eta, 0) = J_m \left(\kappa \sqrt{\xi^2 + \eta^2} \right) e^{-\frac{(\xi^2 + \eta^2)}{2} } e^{i m \varphi},
\label{eqn:BG_Profile}
\end{equation}
where $J_m$ represents the $m^{\mathrm{th}}$ order Bessel function and we choose $\kappa$ such that the size of the central ring of the BG mode matches that of the equivalent LG mode.
BG beams are solutions to the paraxial wave equation that, by controlling the width of the Gaussian, encompass as limiting cases the diffraction-free Bessel beam and the Gaussian beam \cite{GORI1987491,BG_Review}. Typically, these can be made in the lab by utilising a circular slit to transform a plane wave \cite{PhysRevLett.58.1499}, or (specifically for a BG setup) by using an axicon lens to focus a Gaussian beam \cite{garces2002simultaneous}.

As before, we find that $2 |m|$ solitons form. For the weakly {\em repulsive} scattering regime ($\beta_{col} = 3.5$), the diffraction-less nature of the BG beams increases the length that the atoms are confined to the ring to $\approx 1.2 z_R$, and decreases the radial spread of the solitons at $4 z_R$ by $\approx 1.5$ times. If we move to weakly {\em attractive} interactions ($\beta_{col} = -1.5$) we find that, for $m = 1, 2$, the solitons rotate azimuthally with constant velocity along the entire length of the atomic medium, thus producing a form of controllable persistent current.

In conclusion, we have demonstrated the formation of {\em coupled} optical and atomic solitons carrying angular momentum when far-red-detuned light carrying OAM propagates through a BEC.  Despite fundamental differences between the coupled BEC-light model and the pure Kerr case, we find that both optical and atomic fields break into $2 |m|$ solitons as in the Kerr case \cite{PhysRevLett.79.2450,PhysRevLett.87.033901}. These rotate azimuthally around the ring of maximum intensity of the light before breaking away and moving tangentially such that angular momentum is conserved. The number of solitons and their transverse velocity can be controlled by the OAM of the optical beam, with potential applications in atomic transport. By using a Bessel-Gauss beam of equivalent radius and OAM, and moving to weakly attractive interactions we are able to transversely confine the solitons so that they continue to rotate azimuthally for the entire length of the BEC. This has the potential for realising controllable persistent currents in a BEC without the introduction of complex trapping potentials.

\vspace{1ex}
\begin{acknowledgements}
We thank E. Haller for useful discussions. We acknowledge support from EPSRC (EP/R513349/1) via a Doctoral Training Partnership and from the European Training Network ColOpt, which is funded by the European Union (EU) Horizon 2020 program under the Marie Skłodowska-Curie Action, Grant Agreement No. 721465.

\end{acknowledgements}

\bibliography{Refs}

\begin{thebibliography}{34}%
\makeatletter
\providecommand \@ifxundefined [1]{%
 \@ifx{#1\undefined}
}%
\providecommand \@ifnum [1]{%
 \ifnum #1\expandafter \@firstoftwo
 \else \expandafter \@secondoftwo
 \fi
}%
\providecommand \@ifx [1]{%
 \ifx #1\expandafter \@firstoftwo
 \else \expandafter \@secondoftwo
 \fi
}%
\providecommand \natexlab [1]{#1}%
\providecommand \enquote  [1]{``#1''}%
\providecommand \bibnamefont  [1]{#1}%
\providecommand \bibfnamefont [1]{#1}%
\providecommand \citenamefont [1]{#1}%
\providecommand \href@noop [0]{\@secondoftwo}%
\providecommand \href [0]{\begingroup \@sanitize@url \@href}%
\providecommand \@href[1]{\@@startlink{#1}\@@href}%
\providecommand \@@href[1]{\endgroup#1\@@endlink}%
\providecommand \@sanitize@url [0]{\catcode `\\12\catcode `\$12\catcode
  `\&12\catcode `\#12\catcode `\^12\catcode `\_12\catcode `\%12\relax}%
\providecommand \@@startlink[1]{}%
\providecommand \@@endlink[0]{}%
\providecommand \url  [0]{\begingroup\@sanitize@url \@url }%
\providecommand \@url [1]{\endgroup\@href {#1}{\urlprefix }}%
\providecommand \urlprefix  [0]{URL }%
\providecommand \Eprint [0]{\href }%
\providecommand \doibase [0]{http://dx.doi.org/}%
\providecommand \selectlanguage [0]{\@gobble}%
\providecommand \bibinfo  [0]{\@secondoftwo}%
\providecommand \bibfield  [0]{\@secondoftwo}%
\providecommand \translation [1]{[#1]}%
\providecommand \BibitemOpen [0]{}%
\providecommand \bibitemStop [0]{}%
\providecommand \bibitemNoStop [0]{.\EOS\space}%
\providecommand \EOS [0]{\spacefactor3000\relax}%
\providecommand \BibitemShut  [1]{\csname bibitem#1\endcsname}%
\let\auto@bib@innerbib\@empty
\bibitem [{\citenamefont {Mollenauer}\ \emph {et~al.}(1980)\citenamefont
  {Mollenauer}, \citenamefont {Stolen},\ and\ \citenamefont
  {Gordon}}]{PhysRevLett.45.1095}%
  \BibitemOpen
  \bibfield  {author} {\bibinfo {author} {\bibfnamefont {L.~F.}\ \bibnamefont
  {Mollenauer}}, \bibinfo {author} {\bibfnamefont {R.~H.}\ \bibnamefont
  {Stolen}}, \ and\ \bibinfo {author} {\bibfnamefont {J.~P.}\ \bibnamefont
  {Gordon}},\ }\bibfield  {title} {\enquote {\bibinfo {title} {{Experimental
  Observation of Picosecond Pulse Narrowing and Solitons in Optical Fibers}},}\
  }\href {\doibase 10.1103/PhysRevLett.45.1095} {\bibfield  {journal} {\bibinfo
   {journal} {Phys. Rev. Lett.}\ }\textbf {\bibinfo {volume} {45}},\ \bibinfo
  {pages} {1095--1098} (\bibinfo {year} {1980})}\BibitemShut {NoStop}%
\bibitem [{\citenamefont {Wu}\ \emph {et~al.}(1984)\citenamefont {Wu},
  \citenamefont {Keolian},\ and\ \citenamefont
  {Rudnick}}]{PhysRevLett.52.1421}%
  \BibitemOpen
  \bibfield  {author} {\bibinfo {author} {\bibfnamefont {Junru}\ \bibnamefont
  {Wu}}, \bibinfo {author} {\bibfnamefont {Robert}\ \bibnamefont {Keolian}}, \
  and\ \bibinfo {author} {\bibfnamefont {Isadore}\ \bibnamefont {Rudnick}},\
  }\bibfield  {title} {\enquote {\bibinfo {title} {{Observation of a
  Nonpropagating Hydrodynamic Soliton}},}\ }\href {\doibase
  10.1103/PhysRevLett.52.1421} {\bibfield  {journal} {\bibinfo  {journal}
  {Phys. Rev. Lett.}\ }\textbf {\bibinfo {volume} {52}},\ \bibinfo {pages}
  {1421--1424} (\bibinfo {year} {1984})}\BibitemShut {NoStop}%
\bibitem [{\citenamefont {Kosevich}\ \emph {et~al.}(1990)\citenamefont
  {Kosevich}, \citenamefont {Ivanov},\ and\ \citenamefont
  {Kovalev}}]{KOSEVICH1990117}%
  \BibitemOpen
  \bibfield  {author} {\bibinfo {author} {\bibfnamefont {A.M.}\ \bibnamefont
  {Kosevich}}, \bibinfo {author} {\bibfnamefont {B.A.}\ \bibnamefont {Ivanov}},
  \ and\ \bibinfo {author} {\bibfnamefont {A.S.}\ \bibnamefont {Kovalev}},\
  }\bibfield  {title} {\enquote {\bibinfo {title} {{Magnetic Solitons}},}\
  }\href {\doibase https://doi.org/10.1016/0370-1573(90)90130-T} {\bibfield
  {journal} {\bibinfo  {journal} {Physics Reports}\ }\textbf {\bibinfo {volume}
  {194}},\ \bibinfo {pages} {117--238} (\bibinfo {year} {1990})}\BibitemShut
  {NoStop}%
\bibitem [{\citenamefont {Tanaka}(2001)}]{PhysRevLett.88.017002}%
  \BibitemOpen
  \bibfield  {author} {\bibinfo {author} {\bibfnamefont {Y.}~\bibnamefont
  {Tanaka}},\ }\bibfield  {title} {\enquote {\bibinfo {title} {{Soliton in
  Two-Band Superconductor}},}\ }\href {\doibase 10.1103/PhysRevLett.88.017002}
  {\bibfield  {journal} {\bibinfo  {journal} {Phys. Rev. Lett.}\ }\textbf
  {\bibinfo {volume} {88}},\ \bibinfo {pages} {017002} (\bibinfo {year}
  {2001})}\BibitemShut {NoStop}%
\bibitem [{\citenamefont {Dauxois}(2010)}]{Dauxois_2010}%
  \BibitemOpen
  \bibfield  {author} {\bibinfo {author} {\bibfnamefont {Thierry}\ \bibnamefont
  {Dauxois}},\ }\href@noop {} {\emph {\bibinfo {title} {{Physics of
  Solitons}}}}\ (\bibinfo  {publisher} {Cambridge University Press},\ \bibinfo
  {year} {2010})\BibitemShut {NoStop}%
\bibitem [{\citenamefont {Khaykovich}\ \emph {et~al.}(2002)\citenamefont
  {Khaykovich}, \citenamefont {Schreck}, \citenamefont {Ferrari}, \citenamefont
  {Bourdel}, \citenamefont {Cubizolles}, \citenamefont {Carr}, \citenamefont
  {Castin},\ and\ \citenamefont {Salomon}}]{Khaykovich1290}%
  \BibitemOpen
  \bibfield  {author} {\bibinfo {author} {\bibfnamefont {L.}~\bibnamefont
  {Khaykovich}}, \bibinfo {author} {\bibfnamefont {F.}~\bibnamefont {Schreck}},
  \bibinfo {author} {\bibfnamefont {G.}~\bibnamefont {Ferrari}}, \bibinfo
  {author} {\bibfnamefont {T.}~\bibnamefont {Bourdel}}, \bibinfo {author}
  {\bibfnamefont {J.}~\bibnamefont {Cubizolles}}, \bibinfo {author}
  {\bibfnamefont {L.~D.}\ \bibnamefont {Carr}}, \bibinfo {author}
  {\bibfnamefont {Y.}~\bibnamefont {Castin}}, \ and\ \bibinfo {author}
  {\bibfnamefont {C.}~\bibnamefont {Salomon}},\ }\bibfield  {title} {\enquote
  {\bibinfo {title} {{Formation of a Matter-Wave Bright Soliton}},}\ }\href
  {\doibase 10.1126/science.1071021} {\bibfield  {journal} {\bibinfo  {journal}
  {Science}\ }\textbf {\bibinfo {volume} {296}},\ \bibinfo {pages} {1290--1293}
  (\bibinfo {year} {2002})}\BibitemShut {NoStop}%
\bibitem [{\citenamefont {Denschlag}\ \emph {et~al.}(2000)\citenamefont
  {Denschlag}, \citenamefont {Simsarian}, \citenamefont {Feder}, \citenamefont
  {Clark}, \citenamefont {Collins}, \citenamefont {Cubizolles}, \citenamefont
  {Deng}, \citenamefont {Hagley}, \citenamefont {Helmerson}, \citenamefont
  {Reinhardt} \emph {et~al.}}]{denschlag2000generating}%
  \BibitemOpen
  \bibfield  {author} {\bibinfo {author} {\bibfnamefont {J}~\bibnamefont
  {Denschlag}}, \bibinfo {author} {\bibfnamefont {Je~E}\ \bibnamefont
  {Simsarian}}, \bibinfo {author} {\bibfnamefont {Dl~L}\ \bibnamefont {Feder}},
  \bibinfo {author} {\bibfnamefont {Charles~W}\ \bibnamefont {Clark}}, \bibinfo
  {author} {\bibfnamefont {La~A}\ \bibnamefont {Collins}}, \bibinfo {author}
  {\bibfnamefont {J}~\bibnamefont {Cubizolles}}, \bibinfo {author}
  {\bibfnamefont {Lu}~\bibnamefont {Deng}}, \bibinfo {author} {\bibfnamefont
  {Edward~W}\ \bibnamefont {Hagley}}, \bibinfo {author} {\bibfnamefont
  {Kristian}\ \bibnamefont {Helmerson}}, \bibinfo {author} {\bibfnamefont
  {William~P}\ \bibnamefont {Reinhardt}},  \emph {et~al.},\ }\bibfield  {title}
  {\enquote {\bibinfo {title} {{Generating solitons by phase engineering of a
  Bose-Einstein condensate}},}\ }\href {\doibase 10.1126/science.287.5450.97}
  {\bibfield  {journal} {\bibinfo  {journal} {Science}\ }\textbf {\bibinfo
  {volume} {287}},\ \bibinfo {pages} {97--101} (\bibinfo {year}
  {2000})}\BibitemShut {NoStop}%
\bibitem [{\citenamefont {Eiermann}\ \emph {et~al.}(2004)\citenamefont
  {Eiermann}, \citenamefont {Anker}, \citenamefont {Albiez}, \citenamefont
  {Taglieber}, \citenamefont {Treutlein}, \citenamefont {Marzlin},\ and\
  \citenamefont {Oberthaler}}]{PhysRevLett.92.230401}%
  \BibitemOpen
  \bibfield  {author} {\bibinfo {author} {\bibfnamefont {B.}~\bibnamefont
  {Eiermann}}, \bibinfo {author} {\bibfnamefont {Th.}\ \bibnamefont {Anker}},
  \bibinfo {author} {\bibfnamefont {M.}~\bibnamefont {Albiez}}, \bibinfo
  {author} {\bibfnamefont {M.}~\bibnamefont {Taglieber}}, \bibinfo {author}
  {\bibfnamefont {P.}~\bibnamefont {Treutlein}}, \bibinfo {author}
  {\bibfnamefont {K.-P.}\ \bibnamefont {Marzlin}}, \ and\ \bibinfo {author}
  {\bibfnamefont {M.~K.}\ \bibnamefont {Oberthaler}},\ }\bibfield  {title}
  {\enquote {\bibinfo {title} {{Bright Bose-Einstein Gap Solitons of Atoms with
  Repulsive Interaction}},}\ }\href {\doibase 10.1103/PhysRevLett.92.230401}
  {\bibfield  {journal} {\bibinfo  {journal} {Phys. Rev. Lett.}\ }\textbf
  {\bibinfo {volume} {92}},\ \bibinfo {pages} {230401} (\bibinfo {year}
  {2004})}\BibitemShut {NoStop}%
\bibitem [{\citenamefont {Allen}\ \emph {et~al.}(1992)\citenamefont {Allen},
  \citenamefont {Beijersbergen}, \citenamefont {Spreeuw},\ and\ \citenamefont
  {Woerdman}}]{Allen92}%
  \BibitemOpen
  \bibfield  {author} {\bibinfo {author} {\bibfnamefont {L.}~\bibnamefont
  {Allen}}, \bibinfo {author} {\bibfnamefont {M.~W.}\ \bibnamefont
  {Beijersbergen}}, \bibinfo {author} {\bibfnamefont {R.~J.~C.}\ \bibnamefont
  {Spreeuw}}, \ and\ \bibinfo {author} {\bibfnamefont {J.~P.}\ \bibnamefont
  {Woerdman}},\ }\bibfield  {title} {\enquote {\bibinfo {title} {{Orbital
  angular momentum of light and the transformation of Laguerre-Gaussian laser
  modes}},}\ }\href {\doibase 10.1103/PhysRevA.45.8185} {\bibfield  {journal}
  {\bibinfo  {journal} {Phys. Rev. A}\ }\textbf {\bibinfo {volume} {45}},\
  \bibinfo {pages} {8185--8189} (\bibinfo {year} {1992})}\BibitemShut {NoStop}%
\bibitem [{\citenamefont {Firth}\ and\ \citenamefont
  {Skryabin}(1997)}]{PhysRevLett.79.2450}%
  \BibitemOpen
  \bibfield  {author} {\bibinfo {author} {\bibfnamefont {W.~J.}\ \bibnamefont
  {Firth}}\ and\ \bibinfo {author} {\bibfnamefont {D.~V.}\ \bibnamefont
  {Skryabin}},\ }\bibfield  {title} {\enquote {\bibinfo {title} {{Optical
  Solitons Carrying Orbital Angular Momentum}},}\ }\href {\doibase
  10.1103/PhysRevLett.79.2450} {\bibfield  {journal} {\bibinfo  {journal}
  {Phys. Rev. Lett.}\ }\textbf {\bibinfo {volume} {79}},\ \bibinfo {pages}
  {2450--2453} (\bibinfo {year} {1997})}\BibitemShut {NoStop}%
\bibitem [{\citenamefont {Desyatnikov}\ and\ \citenamefont
  {Kivshar}(2001)}]{PhysRevLett.87.033901}%
  \BibitemOpen
  \bibfield  {author} {\bibinfo {author} {\bibfnamefont {Anton~S.}\
  \bibnamefont {Desyatnikov}}\ and\ \bibinfo {author} {\bibfnamefont {Yuri~S.}\
  \bibnamefont {Kivshar}},\ }\bibfield  {title} {\enquote {\bibinfo {title}
  {{Necklace-Ring Vector Solitons}},}\ }\href {\doibase
  10.1103/PhysRevLett.87.033901} {\bibfield  {journal} {\bibinfo  {journal}
  {Phys. Rev. Lett.}\ }\textbf {\bibinfo {volume} {87}},\ \bibinfo {pages}
  {033901} (\bibinfo {year} {2001})}\BibitemShut {NoStop}%
\bibitem [{\citenamefont {Bigelow}\ \emph {et~al.}(2004)\citenamefont
  {Bigelow}, \citenamefont {Zerom},\ and\ \citenamefont {Boyd}}]{Bigelow04}%
  \BibitemOpen
  \bibfield  {author} {\bibinfo {author} {\bibfnamefont {Matthew~S.}\
  \bibnamefont {Bigelow}}, \bibinfo {author} {\bibfnamefont {Petros}\
  \bibnamefont {Zerom}}, \ and\ \bibinfo {author} {\bibfnamefont {Robert~W.}\
  \bibnamefont {Boyd}},\ }\bibfield  {title} {\enquote {\bibinfo {title}
  {{Breakup of Ring Beams Carrying Orbital Angular Momentum in Sodium
  Vapor}},}\ }\href {\doibase 10.1103/PhysRevLett.92.083902} {\bibfield
  {journal} {\bibinfo  {journal} {Phys. Rev. Lett.}\ }\textbf {\bibinfo
  {volume} {92}},\ \bibinfo {pages} {083902} (\bibinfo {year}
  {2004})}\BibitemShut {NoStop}%
\bibitem [{\citenamefont {Bouchard}\ \emph {et~al.}(2016)\citenamefont
  {Bouchard}, \citenamefont {Larocque}, \citenamefont {Yao}, \citenamefont
  {Travis}, \citenamefont {De~Leon}, \citenamefont {Rubano}, \citenamefont
  {Karimi}, \citenamefont {Oppo},\ and\ \citenamefont
  {Boyd}}]{PhysRevLett.117.233903}%
  \BibitemOpen
  \bibfield  {author} {\bibinfo {author} {\bibfnamefont {Fr\'ed\'eric}\
  \bibnamefont {Bouchard}}, \bibinfo {author} {\bibfnamefont {Hugo}\
  \bibnamefont {Larocque}}, \bibinfo {author} {\bibfnamefont {Alison~M.}\
  \bibnamefont {Yao}}, \bibinfo {author} {\bibfnamefont {Christopher}\
  \bibnamefont {Travis}}, \bibinfo {author} {\bibfnamefont {Israel}\
  \bibnamefont {De~Leon}}, \bibinfo {author} {\bibfnamefont {Andrea}\
  \bibnamefont {Rubano}}, \bibinfo {author} {\bibfnamefont {Ebrahim}\
  \bibnamefont {Karimi}}, \bibinfo {author} {\bibfnamefont {Gian-Luca}\
  \bibnamefont {Oppo}}, \ and\ \bibinfo {author} {\bibfnamefont {Robert~W.}\
  \bibnamefont {Boyd}},\ }\bibfield  {title} {\enquote {\bibinfo {title}
  {{Polarization Shaping for Control of Nonlinear Propagation}},}\ }\href
  {\doibase 10.1103/PhysRevLett.117.233903} {\bibfield  {journal} {\bibinfo
  {journal} {Phys. Rev. Lett.}\ }\textbf {\bibinfo {volume} {117}},\ \bibinfo
  {pages} {233903} (\bibinfo {year} {2016})}\BibitemShut {NoStop}%
\bibitem [{\citenamefont {Walasik}\ \emph {et~al.}(2017)\citenamefont
  {Walasik}, \citenamefont {Silahli},\ and\ \citenamefont
  {Litchinitser}}]{Walasik17}%
  \BibitemOpen
  \bibfield  {author} {\bibinfo {author} {\bibfnamefont {W.}~\bibnamefont
  {Walasik}}, \bibinfo {author} {\bibfnamefont {S.~Z.}\ \bibnamefont
  {Silahli}}, \ and\ \bibinfo {author} {\bibfnamefont {N.~M.}\ \bibnamefont
  {Litchinitser}},\ }\bibfield  {title} {\enquote {\bibinfo {title} {{Dynamics
  of necklace beams in nonlinear colloidal suspensions}},}\ }\href {\doibase
  10.1038/s41598-017-12169-x} {\bibfield  {journal} {\bibinfo  {journal} {Sci.
  Rep.}\ }\textbf {\bibinfo {volume} {7}},\ \bibinfo {pages} {11709} (\bibinfo
  {year} {2017})}\BibitemShut {NoStop}%
\bibitem [{\citenamefont {Sun}\ \emph {et~al.}(2018)\citenamefont {Sun},
  \citenamefont {Silahli}, \citenamefont {Walasik}, \citenamefont {Li},
  \citenamefont {Johnson},\ and\ \citenamefont {Litchinitser}}]{Sun18}%
  \BibitemOpen
  \bibfield  {author} {\bibinfo {author} {\bibfnamefont {Jingbo}\ \bibnamefont
  {Sun}}, \bibinfo {author} {\bibfnamefont {Salih~Z.}\ \bibnamefont {Silahli}},
  \bibinfo {author} {\bibfnamefont {Wiktor}\ \bibnamefont {Walasik}}, \bibinfo
  {author} {\bibfnamefont {Qi}~\bibnamefont {Li}}, \bibinfo {author}
  {\bibfnamefont {Eric}\ \bibnamefont {Johnson}}, \ and\ \bibinfo {author}
  {\bibfnamefont {Natalia~M.}\ \bibnamefont {Litchinitser}},\ }\bibfield
  {title} {\enquote {\bibinfo {title} {{Nanoscale orbital angular momentum beam
  instabilities in engineered nonlinear colloidal media}},}\ }\href {\doibase
  10.1364/OE.26.005118} {\bibfield  {journal} {\bibinfo  {journal} {Opt.
  Express}\ }\textbf {\bibinfo {volume} {26}},\ \bibinfo {pages} {5118--5125}
  (\bibinfo {year} {2018})}\BibitemShut {NoStop}%
\bibitem [{\citenamefont {Saffman}\ and\ \citenamefont
  {Skryabin}(2001)}]{Saffman2001}%
  \BibitemOpen
  \bibfield  {author} {\bibinfo {author} {\bibfnamefont {Mark}\ \bibnamefont
  {Saffman}}\ and\ \bibinfo {author} {\bibfnamefont {Dmitry~V.}\ \bibnamefont
  {Skryabin}},\ }\enquote {\bibinfo {title} {{Coupled Propagation of Light and
  Matter Waves: Solitons and Transverse Instabilities}},}\ in\ \href {\doibase
  10.1007/978-3-540-44582-1_17} {\emph {\bibinfo {booktitle} {Spatial
  Solitons}}},\ \bibinfo {editor} {edited by\ \bibinfo {editor} {\bibfnamefont
  {Stefano}\ \bibnamefont {Trillo}}\ and\ \bibinfo {editor} {\bibfnamefont
  {William}\ \bibnamefont {Torruellas}}}\ (\bibinfo  {publisher} {Springer
  Berlin Heidelberg},\ \bibinfo {address} {Berlin, Heidelberg},\ \bibinfo
  {year} {2001})\ pp.\ \bibinfo {pages} {433--447}\BibitemShut {NoStop}%
\bibitem [{\citenamefont {Heckenberg}\ \emph {et~al.}(1992)\citenamefont
  {Heckenberg}, \citenamefont {McDuff}, \citenamefont {Smith},\ and\
  \citenamefont {White}}]{Heckenberg92}%
  \BibitemOpen
  \bibfield  {author} {\bibinfo {author} {\bibfnamefont {N.~R.}\ \bibnamefont
  {Heckenberg}}, \bibinfo {author} {\bibfnamefont {R.}~\bibnamefont {McDuff}},
  \bibinfo {author} {\bibfnamefont {C.~P.}\ \bibnamefont {Smith}}, \ and\
  \bibinfo {author} {\bibfnamefont {A.~G.}\ \bibnamefont {White}},\ }\bibfield
  {title} {\enquote {\bibinfo {title} {{Generation of optical-phase
  singularities by computer generated holograms}},}\ }\href {\doibase
  10.1364/OL.17.000221} {\bibfield  {journal} {\bibinfo  {journal} {Opt.
  Lett.}\ }\textbf {\bibinfo {volume} {17}},\ \bibinfo {pages} {221--223}
  (\bibinfo {year} {1992})}\BibitemShut {NoStop}%
\bibitem [{\citenamefont {Salasnich}\ \emph {et~al.}(2002)\citenamefont
  {Salasnich}, \citenamefont {Parola},\ and\ \citenamefont
  {Reatto}}]{PhysRevA.65.043614}%
  \BibitemOpen
  \bibfield  {author} {\bibinfo {author} {\bibfnamefont {L.}~\bibnamefont
  {Salasnich}}, \bibinfo {author} {\bibfnamefont {A.}~\bibnamefont {Parola}}, \
  and\ \bibinfo {author} {\bibfnamefont {L.}~\bibnamefont {Reatto}},\
  }\bibfield  {title} {\enquote {\bibinfo {title} {{Effective wave equations
  for the dynamics of cigar-shaped and disk-shaped Bose condensates}},}\ }\href
  {\doibase 10.1103/PhysRevA.65.043614} {\bibfield  {journal} {\bibinfo
  {journal} {Phys. Rev. A}\ }\textbf {\bibinfo {volume} {65}},\ \bibinfo
  {pages} {043614} (\bibinfo {year} {2002})}\BibitemShut {NoStop}%
\bibitem [{\citenamefont {Salasnich}\ and\ \citenamefont
  {Malomed}(2009)}]{PhysRevA.79.053620}%
  \BibitemOpen
  \bibfield  {author} {\bibinfo {author} {\bibfnamefont {Luca}\ \bibnamefont
  {Salasnich}}\ and\ \bibinfo {author} {\bibfnamefont {Boris~A.}\ \bibnamefont
  {Malomed}},\ }\bibfield  {title} {\enquote {\bibinfo {title} {{Solitons and
  solitary vortices in pancake-shaped Bose-Einstein condensates}},}\ }\href
  {\doibase 10.1103/PhysRevA.79.053620} {\bibfield  {journal} {\bibinfo
  {journal} {Phys. Rev. A}\ }\textbf {\bibinfo {volume} {79}},\ \bibinfo
  {pages} {053620} (\bibinfo {year} {2009})}\BibitemShut {NoStop}%
\bibitem [{Note1()}]{Note1}%
  \BibitemOpen
  \bibinfo {note} {D. A. Steck, Alkali D Line Data, available online at
  \protect \url {http://steck.us/alkalidata} (2022)}\BibitemShut {NoStop}%
\bibitem [{\citenamefont {Di~Carli}\ \emph {et~al.}(2020)\citenamefont
  {Di~Carli}, \citenamefont {Henderson}, \citenamefont {Flannigan},
  \citenamefont {Colquhoun}, \citenamefont {Mitchell}, \citenamefont {Oppo},
  \citenamefont {Daley}, \citenamefont {Kuhr},\ and\ \citenamefont
  {Haller}}]{PhysRevLett.125.183602}%
  \BibitemOpen
  \bibfield  {author} {\bibinfo {author} {\bibfnamefont {Andrea}\ \bibnamefont
  {Di~Carli}}, \bibinfo {author} {\bibfnamefont {Grant}\ \bibnamefont
  {Henderson}}, \bibinfo {author} {\bibfnamefont {Stuart}\ \bibnamefont
  {Flannigan}}, \bibinfo {author} {\bibfnamefont {Craig~D.}\ \bibnamefont
  {Colquhoun}}, \bibinfo {author} {\bibfnamefont {Matthew}\ \bibnamefont
  {Mitchell}}, \bibinfo {author} {\bibfnamefont {Gian-Luca}\ \bibnamefont
  {Oppo}}, \bibinfo {author} {\bibfnamefont {Andrew~J.}\ \bibnamefont {Daley}},
  \bibinfo {author} {\bibfnamefont {Stefan}\ \bibnamefont {Kuhr}}, \ and\
  \bibinfo {author} {\bibfnamefont {Elmar}\ \bibnamefont {Haller}},\ }\bibfield
   {title} {\enquote {\bibinfo {title} {{Collisionally Inhomogeneous
  Bose-Einstein Condensates with a Linear Interaction Gradient}},}\ }\href
  {\doibase 10.1103/PhysRevLett.125.183602} {\bibfield  {journal} {\bibinfo
  {journal} {Phys. Rev. Lett.}\ }\textbf {\bibinfo {volume} {125}},\ \bibinfo
  {pages} {183602} (\bibinfo {year} {2020})}\BibitemShut {NoStop}%
\bibitem [{\citenamefont {Kraemer}\ \emph {et~al.}(2006)\citenamefont
  {Kraemer}, \citenamefont {Mark}, \citenamefont {Waldburger}, \citenamefont
  {Danzl}, \citenamefont {Chin}, \citenamefont {Engeser}, \citenamefont
  {Lange}, \citenamefont {Pilch}, \citenamefont {Jaakkola}, \citenamefont
  {N{\"a}gerl},\ and\ \citenamefont {Grimm}}]{kraemer2006evidence}%
  \BibitemOpen
  \bibfield  {author} {\bibinfo {author} {\bibfnamefont {Tobias}\ \bibnamefont
  {Kraemer}}, \bibinfo {author} {\bibfnamefont {Manfred}\ \bibnamefont {Mark}},
  \bibinfo {author} {\bibfnamefont {Philipp}\ \bibnamefont {Waldburger}},
  \bibinfo {author} {\bibfnamefont {Johann~G}\ \bibnamefont {Danzl}}, \bibinfo
  {author} {\bibfnamefont {Cheng}\ \bibnamefont {Chin}}, \bibinfo {author}
  {\bibfnamefont {Bastian}\ \bibnamefont {Engeser}}, \bibinfo {author}
  {\bibfnamefont {Almar~D}\ \bibnamefont {Lange}}, \bibinfo {author}
  {\bibfnamefont {Karl}\ \bibnamefont {Pilch}}, \bibinfo {author}
  {\bibfnamefont {Antti}\ \bibnamefont {Jaakkola}}, \bibinfo {author}
  {\bibfnamefont {H-C}\ \bibnamefont {N{\"a}gerl}}, \ and\ \bibinfo {author}
  {\bibfnamefont {R}~\bibnamefont {Grimm}},\ }\bibfield  {title} {\enquote
  {\bibinfo {title} {{Evidence for Efimov quantum states in an ultracold gas of
  caesium atoms}},}\ }\href {\doibase https://doi.org/10.1038/nature04626}
  {\bibfield  {journal} {\bibinfo  {journal} {Nature}\ }\textbf {\bibinfo
  {volume} {440}},\ \bibinfo {pages} {315--318} (\bibinfo {year}
  {2006})}\BibitemShut {NoStop}%
\bibitem [{\citenamefont {Berg\'e}(1998)}]{Berge98}%
  \BibitemOpen
  \bibfield  {author} {\bibinfo {author} {\bibfnamefont {Luc}\ \bibnamefont
  {Berg\'e}},\ }\bibfield  {title} {\enquote {\bibinfo {title} {{Wave collapse
  in physics: principles and applications to light and plasma waves}},}\ }\href
  {\doibase 10.1016/S0370-1573(97)00092-6} {\bibfield  {journal} {\bibinfo
  {journal} {Phys. Rep.}\ }\textbf {\bibinfo {volume} {303}},\ \bibinfo {pages}
  {259--370} (\bibinfo {year} {1998})}\BibitemShut {NoStop}%
\bibitem [{\citenamefont {Carli}\ \emph {et~al.}(2019)\citenamefont {Carli},
  \citenamefont {Colquhoun}, \citenamefont {Kuhr},\ and\ \citenamefont
  {Haller}}]{Carli_2019}%
  \BibitemOpen
  \bibfield  {author} {\bibinfo {author} {\bibfnamefont {A~Di}\ \bibnamefont
  {Carli}}, \bibinfo {author} {\bibfnamefont {C~D}\ \bibnamefont {Colquhoun}},
  \bibinfo {author} {\bibfnamefont {S}~\bibnamefont {Kuhr}}, \ and\ \bibinfo
  {author} {\bibfnamefont {E}~\bibnamefont {Haller}},\ }\bibfield  {title}
  {\enquote {\bibinfo {title} {{Interferometric measurement of micro-g
  acceleration with levitated atoms}},}\ }\href {\doibase
  10.1088/1367-2630/ab1bbd} {\bibfield  {journal} {\bibinfo  {journal} {New
  Journal of Physics}\ }\textbf {\bibinfo {volume} {21}},\ \bibinfo {pages}
  {053028} (\bibinfo {year} {2019})}\BibitemShut {NoStop}%
\bibitem [{\citenamefont {Petrov}\ \emph {et~al.}(2001)\citenamefont {Petrov},
  \citenamefont {Shlyapnikov},\ and\ \citenamefont
  {Walraven}}]{PhysRevLett.87.050404}%
  \BibitemOpen
  \bibfield  {author} {\bibinfo {author} {\bibfnamefont {D.~S.}\ \bibnamefont
  {Petrov}}, \bibinfo {author} {\bibfnamefont {G.~V.}\ \bibnamefont
  {Shlyapnikov}}, \ and\ \bibinfo {author} {\bibfnamefont {J.~T.~M.}\
  \bibnamefont {Walraven}},\ }\bibfield  {title} {\enquote {\bibinfo {title}
  {{Phase-Fluctuating 3D Bose-Einstein Condensates in Elongated Traps}},}\
  }\href {\doibase 10.1103/PhysRevLett.87.050404} {\bibfield  {journal}
  {\bibinfo  {journal} {Phys. Rev. Lett.}\ }\textbf {\bibinfo {volume} {87}},\
  \bibinfo {pages} {050404} (\bibinfo {year} {2001})}\BibitemShut {NoStop}%
\bibitem [{\citenamefont {Barnett}\ and\ \citenamefont
  {Zambrini}(2007)}]{Barnett07}%
  \BibitemOpen
  \bibfield  {author} {\bibinfo {author} {\bibfnamefont {Stephen~M.}\
  \bibnamefont {Barnett}}\ and\ \bibinfo {author} {\bibfnamefont {Roberta}\
  \bibnamefont {Zambrini}},\ }\enquote {\bibinfo {title} {{Orbital Angular
  Momentum of Light}},}\ in\ \href {\doibase 10.1007/0-387-33988-4_12} {\emph
  {\bibinfo {booktitle} {Quantum Imaging}}},\ \bibinfo {editor} {edited by\
  \bibinfo {editor} {\bibfnamefont {Mikhail~I.}\ \bibnamefont {Kolobov}}}\
  (\bibinfo  {publisher} {Springer New York},\ \bibinfo {address} {New York,
  NY},\ \bibinfo {year} {2007})\ pp.\ \bibinfo {pages} {277--311}\BibitemShut
  {NoStop}%
\bibitem [{\citenamefont {Saffman}(1998)}]{PhysRevLett.81.65}%
  \BibitemOpen
  \bibfield  {author} {\bibinfo {author} {\bibfnamefont {M.}~\bibnamefont
  {Saffman}},\ }\bibfield  {title} {\enquote {\bibinfo {title} {{Self-Induced
  Dipole Force and Filamentation Instability of a Matter Wave}},}\ }\href
  {\doibase 10.1103/PhysRevLett.81.65} {\bibfield  {journal} {\bibinfo
  {journal} {Phys. Rev. Lett.}\ }\textbf {\bibinfo {volume} {81}},\ \bibinfo
  {pages} {65--68} (\bibinfo {year} {1998})}\BibitemShut {NoStop}%
\bibitem [{\citenamefont {Donley}\ \emph {et~al.}(2001)\citenamefont {Donley},
  \citenamefont {Claussen}, \citenamefont {Cornish}, \citenamefont {Roberts},
  \citenamefont {Cornell},\ and\ \citenamefont {Wieman}}]{donley2001dynamics}%
  \BibitemOpen
  \bibfield  {author} {\bibinfo {author} {\bibfnamefont {Elizabeth~A}\
  \bibnamefont {Donley}}, \bibinfo {author} {\bibfnamefont {Neil~R}\
  \bibnamefont {Claussen}}, \bibinfo {author} {\bibfnamefont {Simon~L}\
  \bibnamefont {Cornish}}, \bibinfo {author} {\bibfnamefont {Jacob~L}\
  \bibnamefont {Roberts}}, \bibinfo {author} {\bibfnamefont {Eric~A}\
  \bibnamefont {Cornell}}, \ and\ \bibinfo {author} {\bibfnamefont {Carl~E}\
  \bibnamefont {Wieman}},\ }\bibfield  {title} {\enquote {\bibinfo {title}
  {{Dynamics of collapsing and exploding Bose--Einstein condensates}},}\ }\href
  {\doibase 10.1038/35085500} {\bibfield  {journal} {\bibinfo  {journal}
  {Nature}\ }\textbf {\bibinfo {volume} {412}},\ \bibinfo {pages} {295--299}
  (\bibinfo {year} {2001})}\BibitemShut {NoStop}%
\bibitem [{\citenamefont {Cattani}\ \emph {et~al.}(2011)\citenamefont
  {Cattani}, \citenamefont {Kim}, \citenamefont {Anderson},\ and\ \citenamefont
  {Lisak}}]{PhysRevA.83.013608}%
  \BibitemOpen
  \bibfield  {author} {\bibinfo {author} {\bibfnamefont {F.}~\bibnamefont
  {Cattani}}, \bibinfo {author} {\bibfnamefont {A.}~\bibnamefont {Kim}},
  \bibinfo {author} {\bibfnamefont {D.}~\bibnamefont {Anderson}}, \ and\
  \bibinfo {author} {\bibfnamefont {M.}~\bibnamefont {Lisak}},\ }\bibfield
  {title} {\enquote {\bibinfo {title} {{Co-propagating Bose-Einstein
  condensates and electromagnetic radiation: Emission of mutually localized
  structures}},}\ }\href {\doibase 10.1103/PhysRevA.83.013608} {\bibfield
  {journal} {\bibinfo  {journal} {Phys. Rev. A}\ }\textbf {\bibinfo {volume}
  {83}},\ \bibinfo {pages} {013608} (\bibinfo {year} {2011})}\BibitemShut
  {NoStop}%
\bibitem [{\citenamefont {File}\ and\ \citenamefont
  {Mills}(1963)}]{PhysRevLett.10.93}%
  \BibitemOpen
  \bibfield  {author} {\bibinfo {author} {\bibfnamefont {J.}~\bibnamefont
  {File}}\ and\ \bibinfo {author} {\bibfnamefont {R.~G.}\ \bibnamefont
  {Mills}},\ }\bibfield  {title} {\enquote {\bibinfo {title} {{Observation of
  Persistent Current in a Superconducting Solenoid}},}\ }\href {\doibase
  10.1103/PhysRevLett.10.93} {\bibfield  {journal} {\bibinfo  {journal} {Phys.
  Rev. Lett.}\ }\textbf {\bibinfo {volume} {10}},\ \bibinfo {pages} {93--96}
  (\bibinfo {year} {1963})}\BibitemShut {NoStop}%
\bibitem [{\citenamefont {Gori}\ \emph {et~al.}(1987)\citenamefont {Gori},
  \citenamefont {Guattari},\ and\ \citenamefont {Padovani}}]{GORI1987491}%
  \BibitemOpen
  \bibfield  {author} {\bibinfo {author} {\bibfnamefont {F.}~\bibnamefont
  {Gori}}, \bibinfo {author} {\bibfnamefont {G.}~\bibnamefont {Guattari}}, \
  and\ \bibinfo {author} {\bibfnamefont {C.}~\bibnamefont {Padovani}},\
  }\bibfield  {title} {\enquote {\bibinfo {title} {{Bessel-Gauss beams}},}\
  }\href {\doibase https://doi.org/10.1016/0030-4018(87)90276-8} {\bibfield
  {journal} {\bibinfo  {journal} {Optics Communications}\ }\textbf {\bibinfo
  {volume} {64}},\ \bibinfo {pages} {491--495} (\bibinfo {year}
  {1987})}\BibitemShut {NoStop}%
\bibitem [{\citenamefont {McGloin}\ and\ \citenamefont
  {Dholakia}(2005)}]{BG_Review}%
  \BibitemOpen
  \bibfield  {author} {\bibinfo {author} {\bibfnamefont {D}~\bibnamefont
  {McGloin}}\ and\ \bibinfo {author} {\bibfnamefont {K}~\bibnamefont
  {Dholakia}},\ }\bibfield  {title} {\enquote {\bibinfo {title} {{Bessel beams:
  Diffraction in a new light}},}\ }\href {\doibase 10.1080/0010751042000275259}
  {\bibfield  {journal} {\bibinfo  {journal} {Contemporary Physics}\ }\textbf
  {\bibinfo {volume} {46}},\ \bibinfo {pages} {15--28} (\bibinfo {year}
  {2005})}\BibitemShut {NoStop}%
\bibitem [{\citenamefont {Durnin}\ \emph {et~al.}(1987)\citenamefont {Durnin},
  \citenamefont {Miceli},\ and\ \citenamefont {Eberly}}]{PhysRevLett.58.1499}%
  \BibitemOpen
  \bibfield  {author} {\bibinfo {author} {\bibfnamefont {J.}~\bibnamefont
  {Durnin}}, \bibinfo {author} {\bibfnamefont {J.~J.}\ \bibnamefont {Miceli}},
  \ and\ \bibinfo {author} {\bibfnamefont {J.~H.}\ \bibnamefont {Eberly}},\
  }\bibfield  {title} {\enquote {\bibinfo {title} {{Diffraction-free beams}},}\
  }\href {\doibase 10.1103/PhysRevLett.58.1499} {\bibfield  {journal} {\bibinfo
   {journal} {Phys. Rev. Lett.}\ }\textbf {\bibinfo {volume} {58}},\ \bibinfo
  {pages} {1499--1501} (\bibinfo {year} {1987})}\BibitemShut {NoStop}%
\bibitem [{\citenamefont {Garc{\'e}s-Ch{\'a}vez}\ \emph
  {et~al.}(2002)\citenamefont {Garc{\'e}s-Ch{\'a}vez}, \citenamefont {McGloin},
  \citenamefont {Melville}, \citenamefont {Sibbett},\ and\ \citenamefont
  {Dholakia}}]{garces2002simultaneous}%
  \BibitemOpen
  \bibfield  {author} {\bibinfo {author} {\bibfnamefont {V}~\bibnamefont
  {Garc{\'e}s-Ch{\'a}vez}}, \bibinfo {author} {\bibfnamefont {David}\
  \bibnamefont {McGloin}}, \bibinfo {author} {\bibfnamefont {H}~\bibnamefont
  {Melville}}, \bibinfo {author} {\bibfnamefont {Wilson}\ \bibnamefont
  {Sibbett}}, \ and\ \bibinfo {author} {\bibfnamefont {Kishan}\ \bibnamefont
  {Dholakia}},\ }\bibfield  {title} {\enquote {\bibinfo {title} {{Simultaneous
  micromanipulation in multiple planes using a self-reconstructing light
  beam}},}\ }\href {\doibase 10.1038/nature01007} {\bibfield  {journal}
  {\bibinfo  {journal} {Nature}\ }\textbf {\bibinfo {volume} {419}},\ \bibinfo
  {pages} {145--147} (\bibinfo {year} {2002})}\BibitemShut {NoStop}%
\end{thebibliography}%

\end{document}